\documentclass[conference]{IEEEtran}
\IEEEoverridecommandlockouts
\usepackage{cite}
\usepackage{amsmath,amssymb,amsfonts}
\usepackage{algorithmic}
\usepackage{graphicx}
\usepackage{textcomp}
\usepackage{float}
\usepackage{placeins}
\usepackage{subcaption} 
\usepackage{caption}
\usepackage{comment}
\usepackage{lipsum}
\usepackage{url}
\usepackage{xcolor}
\def\BibTeX{{\rm B\kern-.05em{\sc i\kern-.025em b}\kern-.08em
    T\kern-.1667em\lower.7ex\hbox{E}\kern-.125emX}}
\begin{document}
\bstctlcite{IEEEexample:BSTcontrol}

\title{AI-Driven Collaborative Satellite Object Detection for Space Sustainability
\thanks{We acknowledge the support of the Natural Sciences and Engineering Research Council of Canada (NSERC), [funding reference number RGPIN-2022-03364]}
}

\author{\IEEEauthorblockN{
Peng Hu\IEEEauthorrefmark{1}\IEEEauthorrefmark{2} and
Wenxuan Zhang\IEEEauthorrefmark{1}}
\IEEEauthorblockA{\IEEEauthorrefmark{1}Faculty of Mathematics, University of Waterloo, Waterloo, Canada}
\IEEEauthorblockA{\IEEEauthorrefmark{2}
Dept. of Electrical and Computer Engineering, University of Manitoba, Winnipeg, Canada}
\{peng.hu@umanitoba.ca, v39zhang@uwaterloo.ca\}
}

\maketitle

\begin{abstract}
The growing density of satellites in low-Earth orbit (LEO) presents serious challenges to space sustainability, primarily due to the increased risk of in-orbit collisions. Traditional ground-based tracking systems are constrained by latency and coverage limitations, underscoring the need for onboard, vision-based space object detection (SOD) capabilities. In this paper, we propose a novel satellite clustering framework that enables the collaborative execution of deep learning (DL)-based SOD tasks across multiple satellites. To support this approach, we construct a high-fidelity dataset simulating imaging scenarios for clustered satellite formations. A distance-aware viewpoint selection strategy is introduced to optimize detection performance, and recent DL models are used for evaluation. Experimental results show that the clustering-based method achieves competitive detection accuracy compared to single-satellite and existing approaches, while maintaining a low size, weight, and power (SWaP) footprint. These findings underscore the potential of distributed, AI-enabled in-orbit systems to enhance space situational awareness and contribute to long-term space sustainability.
\end{abstract}


\section{Introduction}
\IEEEPARstart{T}{he} rapid expansion of advanced low-Earth-orbit (LEO) satellite mega-constellations is reshaping the future of space operations, promising global Internet coverage and enhanced near-Earth communication infrastructure. However, managing thousands of these satellites while ensuring space safety and long-term sustainability presents a growing challenge \cite{Hu_SatAIOps}. With a highly dynamic space environment, the presence of countless space objects, unpredictable atmospheric conditions, and unforeseen anomalies such as orbital shifts, power failures, or satellite malfunctions, the risk of collisions becomes significant, threatening both the benefits of LEO satellite systems and the broader space environment. 

The risk of in-orbit collisions is exacerbated by increasingly dense satellite deployments, raising concerns about the Kessler syndrome. Real-world incidents underscore this danger. For example, a collision incident occurred between the Iridium and Kosmos satellites in 2009 \cite{Nicholas2009}, and more recently, nearly 25,000 collision-avoidance maneuvers were conducted between December 1, 2022, and May 31, 2023. Even under highly optimistic conditions, such as high compliance with debris mitigation guidelines and no in-orbit explosions, collisions are still expected to occur every 5 to 9 years \cite{ESA_cleanspace}.

Active monitoring of nearby space objects is a key strategy for space object detection (SOD) that facilitates detecting and assessing collision risks. Traditional ground-based radar and optical tracking systems have been used for this purpose, but they fall short in scenarios requiring real-time response. In-orbit sensing has emerged as a viable alternative, with common modalities including radar, LiDAR, and vision. However, due to limitations in detection range and strict size, weight, and power (SWaP) constraints, radar and LiDAR systems are often impractical for small satellites \cite{zhang2024sensing}. Vision sensors, by contrast, offer a compelling solution, yet few have addressed their potential for tackling emerging space sustainability challenges.

Recent studies have shown that vision-based onboard SOD systems \cite{Hu_SatAIOps, zhang2024sensing, sodv2} can be both effective and energy-efficient, even on embedded GPUs. For instance, our recently proposed deep learning (DL) models, such as GELAN-ViT and GELAN-ViT-SE, achieve peak power consumption around 2000 mW and average power near 1750 mW, outperforming the typical GELAN-t model in YOLOv9 \cite{7780460}. While improvements in precision are still needed, especially for onboard deployment, these methods are notable for their ability to run using only the resources of a single satellite.

Given the growing number of LEO satellites and improved onboard computing capabilities, two critical questions arise: \textit{Can we leverage multiple LEO satellites to perform SOD collaboratively? And how effective is this multi-satellite approach?} In this paper, we address these questions through an initial exploration into collaborative SOD using multiple satellites. As the first to investigate this direction, we introduce the concept of satellite clustering to enable cooperative execution of SOD tasks. We then present an AI-driven solution leveraging DL models, demonstrating its effectiveness compared to recently proposed approaches. Our key contributions are as follows:

\begin{itemize}
\item We propose a satellite clustering framework that enables distributed execution of DL models for SOD tasks.
\item We present a novel dataset comprising high-fidelity satellite assets to support AI-driven satellite clustering solutions.
\item We evaluate the proposed clustering solution in comparison to single-satellite baselines with recent DL–based models for SOD.

\end{itemize}

The remainder of the paper is organized as follows: Section II reviews related work; Section III introduces the system model and the proposed solution; Section IV presents the experimental results; and Section V concludes the paper and outlines directions for future work.

\section{Related Work}
The space-based visible (SBV) sensors \cite{SBV_01} have long supported routine surveillance of resident space objects (RSOs) in the geosynchronous belt. These systems use onboard signal processing to analyze focal-plane imagery, though processing four consecutive frame sets can take up to 200 seconds. Originally focused on the geosynchronous region, SBV sensors have been extended to observe objects in the geostationary transfer orbit (GTO) and medium Earth orbit (MEO) \cite{HU20171751}. Beyond surveillance, computer vision plays a vital role in space missions such as autonomous navigation, path planning around non-cooperative targets \cite{Mahendrakar21}, and robotic docking or object capture. In close-range rendezvous operations, robotic arms rely on vision systems to locate targets. However, current SBV solutions are inefficient and do not meet the stringent performance requirements for SOD tasks. 

Most modern DL models in computer vision rely on Convolutional Neural Networks (CNNs), which are particularly well-suited for extracting local features such as edges, textures, and object shapes from the images captured by vision sensor payloads for SOD tasks. Among the most widely used CNN-based architectures are Faster R-CNN \cite{NIPS2015_14bfa6bb} and the You Only Look Once (YOLO) family of models \cite{7780460}. Faster R-CNN addresses the timing inefficiency by computing a shared convolutional feature map for the entire image, significantly reducing processing time. In contrast, YOLO adopts a single-stage approach by dividing the image into grids, where each grid predicts multiple bounding boxes along with their class probabilities, enabling real-time object detection. The YOLO family, in particular, has been favored in real-time applications because of its single-stage design that balances inference speed and detection precision. YOLOv9 \cite{wang2024yolov9learningwantlearn}, one of the state-of-the-art models in the YOLO series, introduces the lightweight Generalized Efficient Layer Aggregation Network (GELAN) architecture to improve the aggregation of features across layers.

However, CNN-based detectors often struggle with small object detection, which is a key challenge in SOD. Objects of interest in orbital imagery often occupy only a few pixels due to the high orbital altitude of imaging satellites and the vast field of view they cover. The limited resolution of the images also reduces the discriminative features available for detection and increases the difficulty of distinguishing true objects from background noise. Further, introduced by Dosovitskiy \textit{et al.} \cite{Dosovitskiy2020AnII}, the vision transformer (ViT)-based detectors have been explored as an alternative to CNN-based models. Their ability to capture long-range dependencies provides an advantage in detecting small objects; however, their high computational complexity limits their applicability in real-time onboard environments. In our prior work \cite{sodv2}, we proposed GELAN-ViT-SE, a hybrid model that combines CNN and ViT to enhance detection performance in SOD.

Although recent DL models have demonstrated the ability to meet in-situ performance and efficiency requirements, they are limited to single-satellite scenarios. This paper explores the collaborative use of multiple LEO satellites employing DL-based models for SOD, with GELAN-t and GELAN-ViT-SE serving as representative architectures.



\section{System Model \& Proposed Solution}
We first discuss the satellite clustering and then the key topics associated, such as viewpoints, distances from satellites to space objects being monitored, and pairwise distances between satellites. We also discuss the custom dataset we generated to facilitate our study of these concepts.

\subsection{Satellite Cluster Modeling}
We define a satellite cluster as a set of proximity satellites whose fields of view cover a set of objects of interest. Formally, the $i$-th cluster is denoted by:

\[
S(i) = \{s_{i,1}, s_{i,2}, \ldots, s_{i,k}\}
\]

Each satellite $s_{i,j}$ has an associated camera viewing angle $a_{i,j}$, and the set of all such angles in cluster $i$ is:

\[
a(i) = \{a_{i,1}, a_{i,2}, \ldots, a_{i,k}\}
\]

The objects of interest for cluster $i$ are represented as:

\[
d(i) = \{d_{i,1}, d_{i,2}, \ldots, d_{i,m}\}
\]

For a given satellite $s_{i,j}$, we define the objects in the view as:

\[
d(i, s_{i,j}) \subseteq d(i)
\]

For simplicity, we assume that the objects in $d(i, s_{i,j})$ are sorted in ascending order of distance from the satellite. That is, $d(i, s_{i,j})_1$ represents the closest object visible to $s_{i,j}$.

We select $s_{i,1}$ as the central satellite of each cluster, and every satellite $s_{i,j}$ in the cluster is required to satisfy two conditions: it must lie within a distance $r_i$ from the central satellite $s_{i,1}$, and it must observe at least one object of interest such that $d(i, s_{i,j}) \neq \emptyset$.

We further define viewpoint $V_j$ as the collection of all $j$-th satellites across clusters:

\[
V_j = \{ s_{i,j} \mid \forall i \}
\]

With the defined satellite clustering, various collaborative schemes can be employed. In our focused discussion, we consider only the essential communication among satellites within a cluster, where satellites exchange short messages via available inter-satellite links (ISLs), which can support the viewpoint selection strategy. Given that ISLs, particularly free-space optical (FSO) ISLs, can support data rates as high as 1--10 Gbps, the resulting communication overhead is minimal.

\subsection{Distance-Based Viewpoint Selection Strategy}

Given that a target's apparent area on a sensor (in pixels) intuitively decreases approximately with the square of its distance, we propose a distance-based viewpoint selection strategy to enhance object detection performance across satellite clusters. For each cluster $S(i) = \{s_{i,1}, s_{i,2}, s_{i,3}\}$, the strategy selects a single satellite $s_{i,j^*}$ per cluster, where the selected satellite minimizes the average distance to its visible targets $d(i, s_{i,j})$:

\[
s_{i,j^*} = \arg\min_{j} 
\left( 
\frac{1}{|d(i, s_{i,j})|} 
\sum_{d_k \in d(i, s_{i,j})} 
\|s_{i,j} - d_k\|
\right)
\]

The resulting set of selected satellites, one per cluster, is denoted by $V_d$.

\subsection{Discussion of Other Selection Strategy}
While collaborative object detection often involves multi-view fusion techniques, our approach intentionally selects a single, optimal view from the satellite cluster. We explored several multi-view fusion methods, such as voting, bounding box merging, and early fusion, but found them to be unsuitable for our scenario.

\subsubsection{Bounding Box Merging and Voting}
Methods like bounding box merging and voting aim to improve detection accuracy by combining results from multiple views. In theory, if several satellites detect the same object, their individual bounding boxes can be merged or averaged to produce a single, more precise detection.

However, this approach relies on the ability to perfectly align the different viewpoints, which is challenging in the satellite cluster scenario. The position differences between satellites in a cluster result in large differences in viewing angles and perspectives. This challenge is further exacerbated by the small size of the space objects, where even minor geometric misalignment can lead to registration failure. Simple geometric transformations are insufficient to align detections across these distinct viewpoints, leading to high registration errors. In our experiments, directly merging or voting on bounding boxes from different satellite views led to a degradation in detection results compared to our single-view selection strategy, particularly for large clusters with bigger distance spread.

\subsubsection{Early Fusion}

Early fusion is a technique where data from multiple sources is combined into a single input before being processed by a model. In this scenario, images from different satellites would be stacked together, allowing the model to analyze all viewpoints simultaneously. This approach could enhance detection robustness, as the model can learn complex patterns and correlations directly from the combined multi-view data.

However, this strategy is severely constrained by its immense communication overhead. To create the combined input, one satellite would need to receive the full imagery from all other participating satellites for every single frame. Transferring this volume of data in real-time is currently prohibitive given the bandwidth limitations of inter-satellite communication links.

\subsection{Dataset Generation}

To facilitate the proposed research, we develop a satellite clustering for SOD (SCS) dataset\footnote{Available:  \url{https://github.com/AEL-Lab/SOD-Clustering}} that can simulate scenarios involving multiple nearby satellites capturing the same scene from slightly different positions at identical viewing angles. The goal of this dataset is to study how satellite clusters influence SOD. 

\subsection{Framework}

The SCS dataset is generated using Unity, which simulates a realistic solar system that represents real-world physics for celestial body and LEO satellite dynamics. In this simulation, LEO satellites are initially spawned at altitudes randomly chosen between 500 km and 600 km above Earth's surface, with random orbital placements.

The simulation replicates an onboard satellite camera with a fixed Field of View (FOV). To capture data, the onboard camera is attached to a specific satellite, capturing the surrounding environment with a fixed 45-degree FOV. The simulation script identifies the nearest satellite and adjusts the camera's orientation to ensure this target satellite is centrally captured. The script then records metadata alongside each image, including the distances between the camera and all visible satellites. This process continues iteratively, with the onboard camera transitioning to another satellite within the simulation after an image is captured from one satellite's perspective, systematically visiting all satellites in a batch. Upon completion of a batch (comprising 1,000 satellites), a new batch is generated, and the image capture cycle repeats until the desired dataset size is achieved.

\subsection{Dataset}

\begin{figure}\centering

\begin{subfigure}[t]{0.15\textwidth}
    \includegraphics[width=\linewidth]{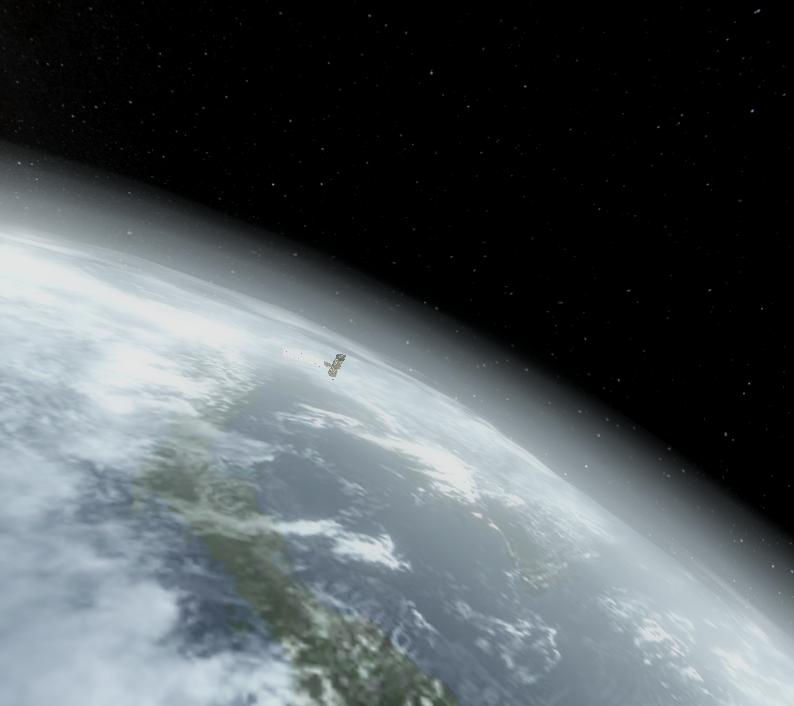}
    \caption{Close - $s_{i,1}$}
\end{subfigure}
\begin{subfigure}[t]{0.15\textwidth}
    \includegraphics[width=\linewidth]{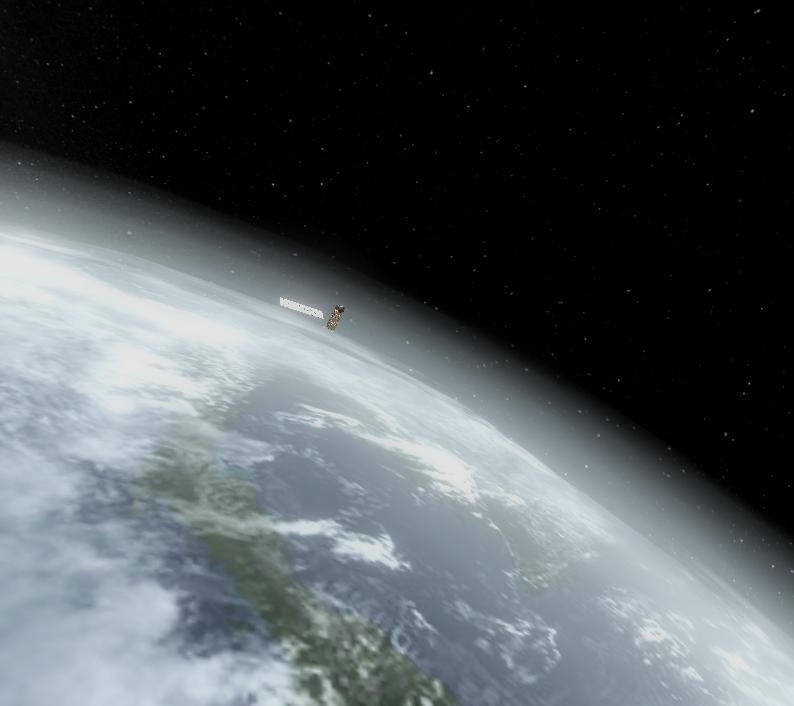}
    \caption{Close - $s_{i,2}$}
\end{subfigure}
\begin{subfigure}[t]{0.15\textwidth}
    \includegraphics[width=\linewidth]{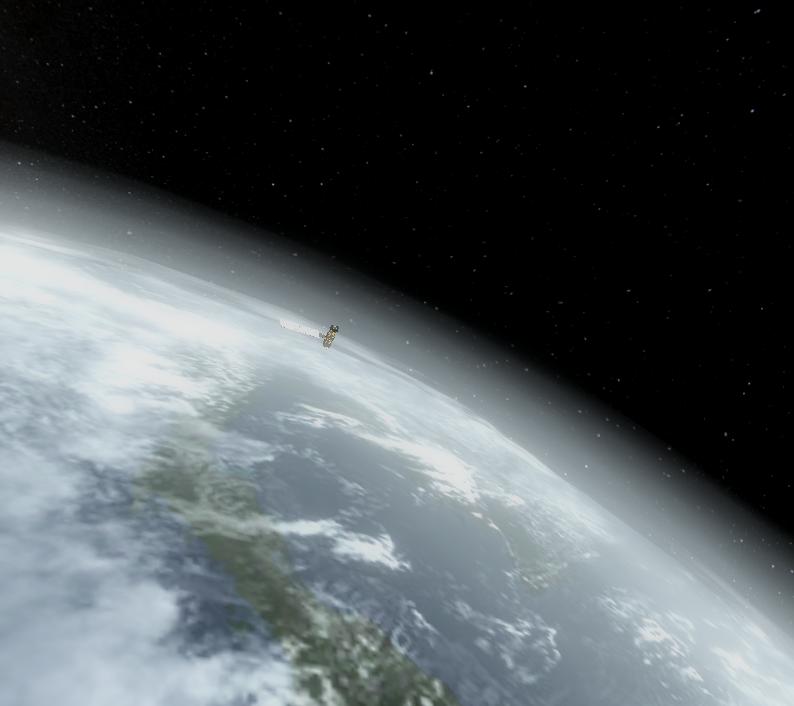}
    \caption{Close - $s_{i,3}$}
\end{subfigure}

\begin{subfigure}[t]{0.15\textwidth}
    \includegraphics[width=\linewidth]{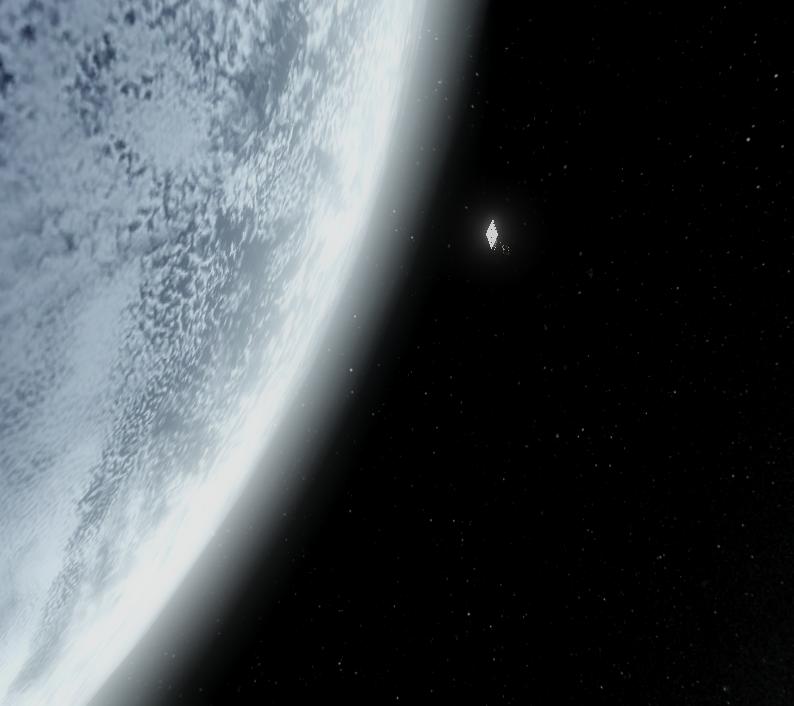}
    \caption{Mid - $s_{i,1}$}
\end{subfigure}
\begin{subfigure}[t]{0.15\textwidth}
    \includegraphics[width=\linewidth]{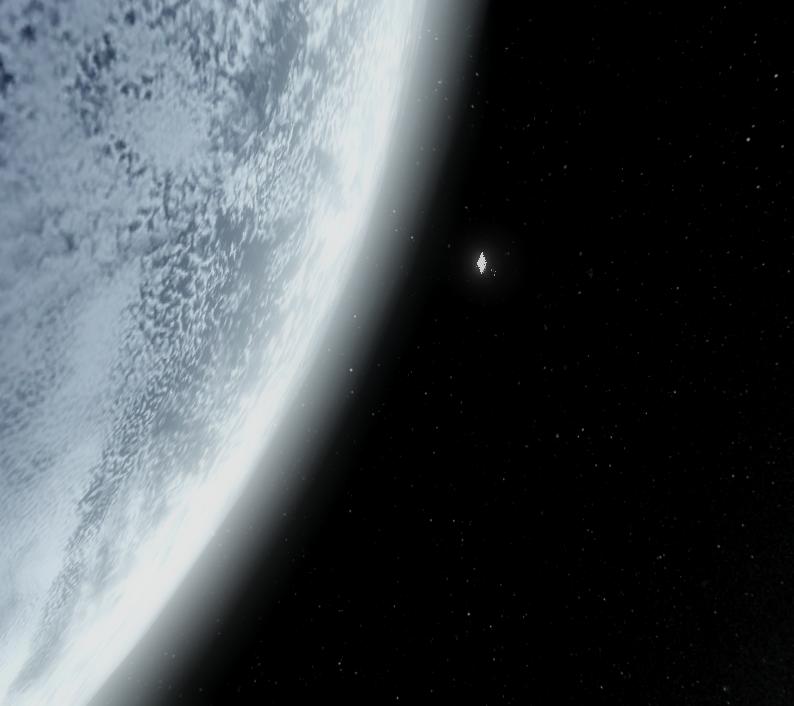}
    \caption{Mid - $s_{i,2}$}
\end{subfigure}
\begin{subfigure}[t]{0.15\textwidth}
    \includegraphics[width=\linewidth]{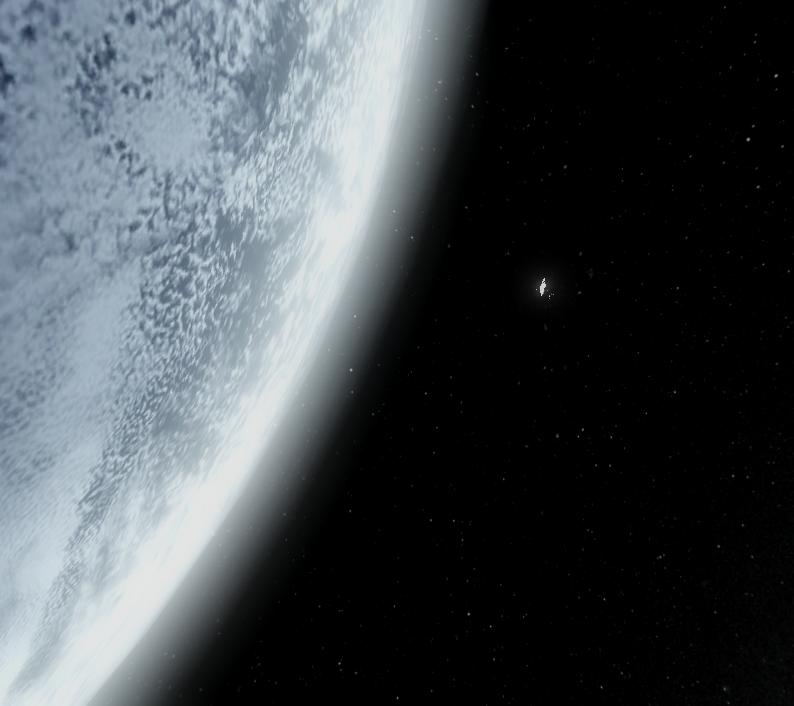}
    \caption{Mid - $s_{i,3}$}
\end{subfigure}

\begin{subfigure}[t]{0.15\textwidth}
    \includegraphics[width=\linewidth]{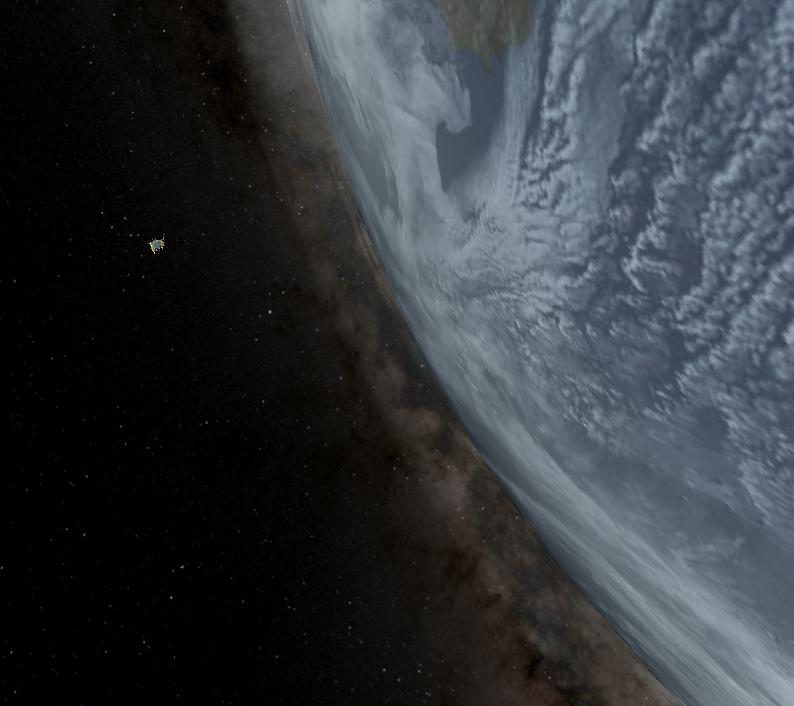}
    \caption{Far - $s_{i,1}$}
\end{subfigure}
\begin{subfigure}[t]{0.15\textwidth}
    \includegraphics[width=\linewidth]{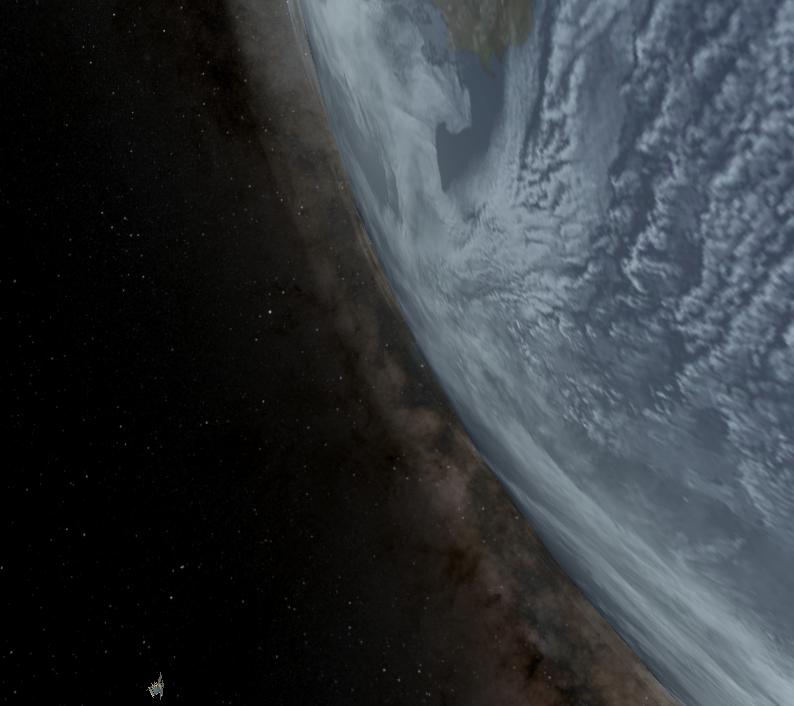}
    \caption{Far - $s_{i,2}$}
\end{subfigure}
\begin{subfigure}[t]{0.15\textwidth}
    \includegraphics[width=\linewidth]{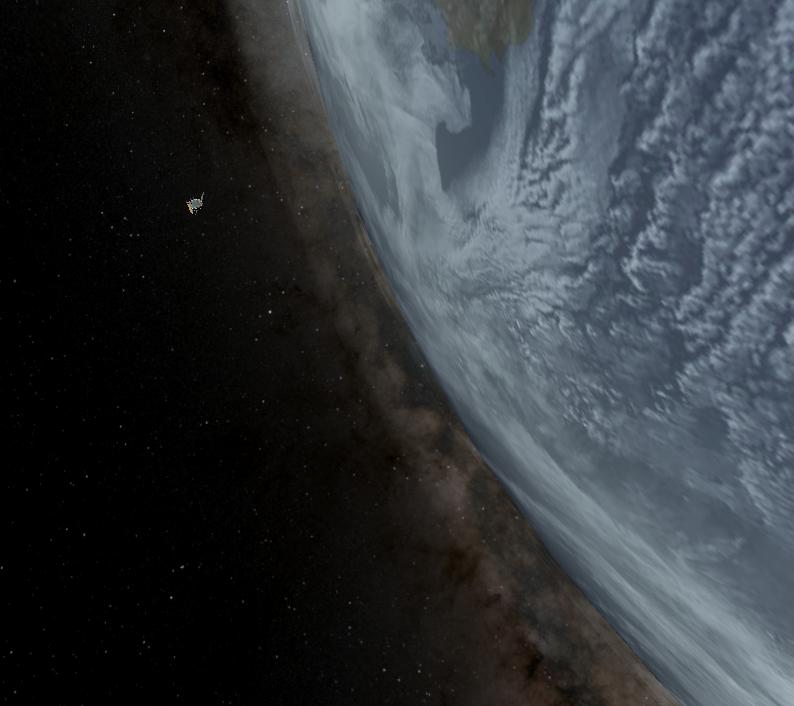}
    \caption{Far - $s_{i,3}$}
\end{subfigure}

\caption{Sample views across three satellite cluster types: Close, Mid, and Far.}
\label{fig:3x3_fullwidth}
\end{figure}

In SCS dataset, a cluster is defined as a set $S(i) = \{s_{i,1}, s_{i,2}, s_{i,3}\}$, with $s_{i,1}$ being the central satellite, and $s_{i,2}$ and $s_{i,3}$ being the secondary satellites within a defined radius $r_i$ of $s_{i,1}$. The example views from the dataset can be seen from Fig. \ref{fig:3x3_fullwidth}.

Here we adopt a cluster size of $k=3$ as it offers a sufficient and representative spatial offset between viewpoints for analyzing satellite clusters. Although our approach can naturally extend to larger clusters, we find that three viewpoints effectively capture the key benefits of spatial diversity. From another perspective, the chosen cluster size aligns well with the common assumption of having four neighboring satellites available across intra-plane and inter-plane configurations. While larger clusters are possible, this assumption helps reduce operational complexity and the overhead associated with coordinating additional satellites.

In each cluster, the central satellite $s_{i,1}$ observes its closest object of interest, $d(i, s_{i,1})_1$, at distances between 0.5 km and 2 km from the camera. The secondary satellites $s_{i,2}$ and $s_{i,3}$ are generated such that their positions lie within the radius range from the central satellite. Specifically, close clusters are defined by a radius of 0.5 km, mid clusters by 1 km, and far clusters by 2 km. The positions of the secondary satellites are sampled uniformly within these distance constraints while maintaining identical viewing angles. These distance ranges are considered based on LEO object speeds. Our recent studies \cite{zhang2024sensing, sodv2} show an inference time of $30-60$ ms is sufficient for detecting objects within close proximity (less than $0.5$ km), meeting the requirement of $\leq 64.1$ ms when considering an average LEO-object speed of $7.8$ km/s. 

Each cluster consists of three satellites capturing images from similar viewpoints. This results in three distinct images of the same scene, taken from slightly different positions but with the same viewing angle. Each cluster contains 60 images (comprising 20 scenes, with three images per scene) leading to a total of 180 images in the dataset.
\begin{table}[ht]
\centering
\caption{Average distance (km) from each image to the objects of interest within view, grouped by viewpoint ($V(1)$, $V(2)$, $V(3)$, and $V_d$) and cluster type (close, mid, far).}
\begin{tabular}{|l|c|c|c|c|}
\hline
Cluster & $V(1)$ & $V(2)$ & $V(3)$ & $V_d$\\
\hline
Close   & 2.98  & 2.917  & 2.86 & 2.80\\
Mid     & 2.57  & 2.89  & 2.63 & 2.37\\
Far     & 2.16  & 2.60  & 2.70 & 2.03\\
Overall & 2.57  & 2.80  & 2.73  & 2.40\\
\hline
\end{tabular}
\label{tab:distance_by_cluster}
\end{table}

To provide a comprehensive characterization of the spatial properties within our SCS dataset, we analyze the average distances between our simulated satellites and their visible targets. Table \ref{tab:distance_by_cluster} presents these average distances, derived directly from the metadata collected during the simulations.

\subsection{Distance from Satellites to Objects of Interest}

\begin{figure*}[ht]
    \centering
    \includegraphics[width=0.69\linewidth]{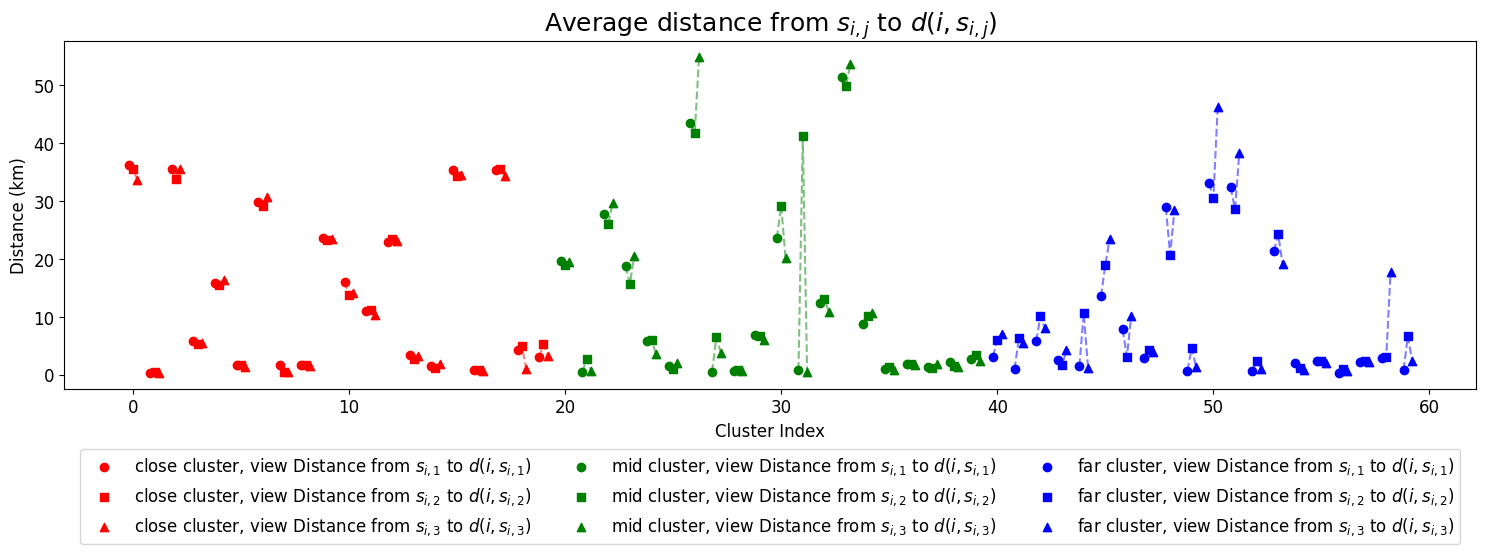}
    \caption{Average distance from each satellite to all objects of interest in view.}
    \label{fig:distance_to_satellite}

\end{figure*}

\begin{figure}[t!]
    \centering
    \includegraphics[width=\linewidth]{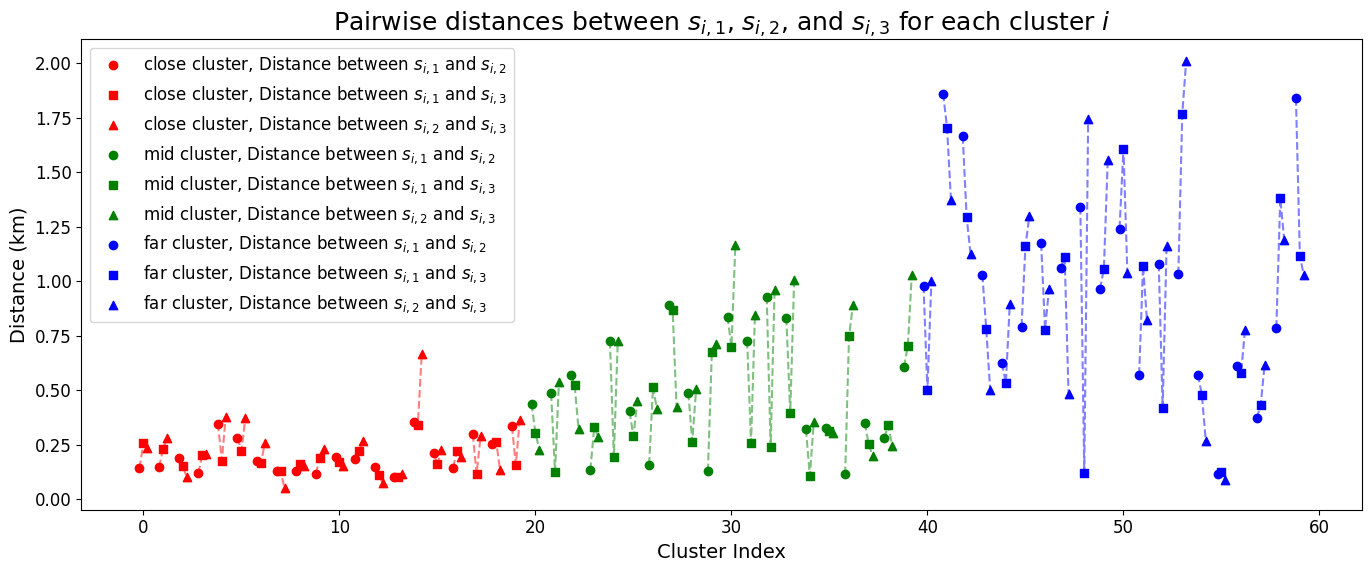}
    \caption{Pairwise distances between different $s_{i, 1}$, $s_{i, 2}$, $s_{i, 3}$ across all clusters.}
    \label{fig:viewpoint_distances}
\end{figure}

Satellites within the same cluster tend to exhibit similar average distances to their visible targets. However, many of these average distances exceed the 0.5–2.0 km initialization range. This occurs because only the nearest object to the central satellite $s_{i,1}$ is constrained to lie within that range; when multiple objects are present, the remaining objects can be positioned farther away, thereby increasing the mean distance.

Table \ref{tab:distance_by_cluster} shows that the central satellite $V(1)$ tends to have the smallest average distance to the object set compared to $V(2)$ and $V(3)$. This is because the secondary satellites are sampled from a sphere centered on the primary satellite. The probability that a secondary satellite is closer to an object's center, $O$, than the primary satellite is less than half. This is due to the geometric fact that the sampled region where the secondary satellite could be closer to $O$ forms a spherical cap that is smaller than a hemisphere of the sampling sphere. This constraint makes it more likely for the secondary viewpoints to have a greater average distance to the objects of interest. In the SCS database, the average distance can be seen in Fig. \ref{fig:distance_to_satellite}.

\subsection{Pairwise Distances Between Satellites}

Fig. \ref{fig:viewpoint_distances} presents the average pairwise distances between satellites $s_{i,1}$, $s_{i,2}$, and $s_{i,3}$ across all clusters. The overall trend follows the design of our clustering strategy: the average pairwise distances increase in the order of close, mid, and far clusters.

Each secondary satellite $s_{i,2}$ and $s_{i,3}$ is sampled within a solid sphere of radius $r$ centered at the primary satellite $s_{i,1}$. As a result, the distances $\|s_{i,1} - s_{i,2}\|$ and $\|s_{i,1} - s_{i,3}\|$ are guaranteed to remain within $r$. However, the distance between the two secondary satellites, $\|s_{i,2} - s_{i,3}\|$, is not directly constrained. Since both are independently sampled within the same radius around $s_{i,1}$, their separation can reach up to $2r$ in the worst-case scenario. This explains the wider spread observed in the $\|s_{i,2}$–$s_{i,3}\|$ compared to the other two pairings.

\subsection{Geographic Distribution of Clusters}
\begin{figure}[ht]
    \centering
    \includegraphics[width=1\linewidth]{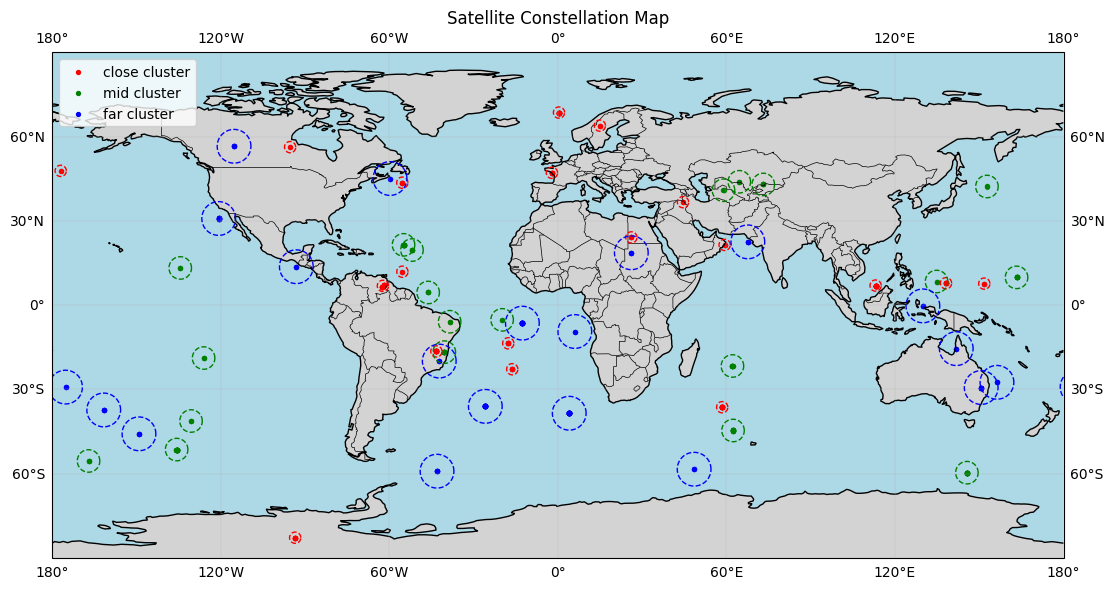}
    \caption{Satellite cluster plotted on a world map.}
    \label{fig:constellation_map}
    \vspace{-12pt}

\end{figure}

In addition to image data, the dataset records detailed metadata for satellites and the objects of interest within view. For each satellite that captures an image, the metadata includes its latitude, longitude, altitude, orbital inclination, and right ascension of the ascending node (RAAN). For each space object of interest visible in an image, the dataset provides the object ID, screen position, bounding box annotation, distance from the observing satellite, geographic coordinates (latitude, longitude, altitude), and orbital parameters (orbital inclination and RAAN).

As an example, Fig. \ref{fig:constellation_map} illustrates the distribution of all recorded satellite cluster locations. The dashed circles indicate cluster sizes based on spread in degrees of latitude and longitude. The circles in Fig. \ref{fig:constellation_map} are for visualization and do not represent the actual distances between satellites within each cluster.

\section{Performance Evaluation}
\begin{table}[ht]
\centering
\caption{Model parameter for GELAN-t and GELAN-ViT-SE}
\begin{tabular}{|l|c|c|c|}
\hline
Model & Inference Time & GFLOPs & Parameter Count\\
\hline
GELAN-t  & 45.14  & 7.3 & 1,913,443\\
GELAN-ViT-SE   & 56.47 & 5.6 & 7,362,012\\
\hline
\end{tabular}
\label{tab:model_parameter}
\end{table}

We conducted all experiments using the typical GELAN-t model \cite{zhang2024sensing} and the latest GELAN-ViT-SE model \cite{sodv2}. To provide context for the models utilized in this study, Table \ref{tab:model_parameter} presents key performance parameters, including inference time, GFLOPs, and parameter count. 

We have trained both models on the SODv2 dataset\footnote{Available: https://github.com/AEL-Lab/satellite-object-detection-dataset-v2} for 1000 epochs with a batch size of 16 \cite{sodv2}. 
For evaluation, we used a single trained model to perform inference on the SCS dataset. Specifically, we evaluated the model’s performance across four data partitions derived from the SCS dataset: the three fixed-viewpoint subsets $V(1)$, $V(2)$, $V(3)$, and the distance-based selection subset $V_d$. These subsets represent potential strategies for viewpoint selection in a satellite clustering-based SOD solution. The metrics used are the mean Average Precision at an Intersection over Union (IoU) threshold of 0.5 (mAP50), and at multiple IoU thresholds ranging from 0.5 to 0.95 (mAP50:95).

All experiments were conducted on Google Colab using an NVIDIA T4 GPU. Embedded GPUs were not used in this study, as the models deployed in the satellite clusters have already been validated on Jetson Orin Nano boards for accuracy, memory usage, and power consumption \cite{sodv2}.

\subsection{Results}

\begin{table}[ht]
\centering
\caption{Detection performance with GELAN-t across distance-based clusters.}
\resizebox{0.5\textwidth}{!}{%
\begin{tabular}{|c|cc|cc|cc|cc|}
\hline
Viewpoint  & \multicolumn{2}{c|}{Close Cluster} & \multicolumn{2}{c|}{Mid Cluster} & \multicolumn{2}{c|}{Far Cluster} & \multicolumn{2}{c|}{Overall} \\
\cline{2-9}
 & mAP50 & mAP50:95 & mAP50 & mAP50:95 & mAP50 & mAP50:95 & mAP50 & mAP50:95 \\
\hline
$V(1)$  & 0.484 & 0.191 & 0.616 & 0.204 & 0.616 & 0.240 & 0.563 & 0.206\\
$V(2)$  & 0.570  & 0.192 & 0.456 & 0.175 & 0.541 & 0.186 & 0.517 & 0.177\\
$V(3)$  & 0.483 & 0.196 & 0.587 & 0.207 & 0.539 & 0.201 & 0.536 & 0.195\\
$V_d$  & 0.527 & 0.173 & 0.526 & 0.170 & 0.712 & 0.249 & 0.579 & 0.187\\
\hline
\end{tabular}%
}
\label{tab:performance_by_cluster}
\vspace{-9pt}
\end{table}

\begin{table}[ht]
\centering
\caption{Detection performance with GELAN-ViT-SE across distance-based clusters.}
\resizebox{0.5\textwidth}{!}{
\begin{tabular}{|c|cc|cc|cc|cc|}
\hline
Viewpoint & \multicolumn{2}{c|}{Close Cluster} & \multicolumn{2}{c|}{Mid Cluster} & \multicolumn{2}{c|}{Far Cluster} & \multicolumn{2}{c|}{Overall} \\
\cline{2-9}
          & mAP50 & mAP50:95 & mAP50 & mAP50:95 & mAP50 & mAP50:95 & mAP50 & mAP50:95 \\
\hline
$V(1)$    & 0.610 & 0.217    & 0.568 & 0.176    & 0.674 & 0.256    & 0.611 & 0.207 \\
$V(2)$    & 0.596 & 0.220    & 0.576 & 0.198    & 0.536 & 0.189    & 0.565 & 0.197 \\
$V(3)$    & 0.629 & 0.242    & 0.473 & 0.157    & 0.638 & 0.247    & 0.569 & 0.208 \\
$V_d$     & 0.600 & 0.185    & 0.558 & 0.166    & 0.708 & 0.235    & 0.615 & 0.190 \\
\hline
\end{tabular}
}
\label{tab:ViT_SE_performance}
\vspace{-12pt}
\end{table}

Table \ref{tab:performance_by_cluster} presents the results of the proposed satellite clustering solution using the GELAN-t model, averaged across distance-based clusters (close, mid, far, and overall). In terms of mAP50, $V_d$ led to an overall improvement; and the most significant gains were observed in the far clusters, where the $V_d$ consistently outperformed all individual viewpoints. In contrast, improvements in the close and mid clusters were more modest. In these cases, $V_d$ performance closely approximated the midpoint between the highest and lowest performing individual viewpoints. Specifically, in the close cluster, $V_d$ achieved an mAP50 of 0.527, nearly equal to the midpoint of 0.5265 between $V(2)$ and $V(3)$. In the mid cluster, $V_d$ score was 0.526, slightly below the midpoint of 0.536 defined by $V(1)$ and $V(2)$.

For mAP50:95, $V_d$ led to improvements in the far clusters but showed a slight decrease in the close and mid clusters, suggesting a potential trade-off in fine-grained localization accuracy.  This trade-off stems directly from the definition of the mAP50:95 metric, which averages scores across a range of IoU thresholds from 0.50 to 0.95. The metric's stringency at high IoU values inherently favors the precise localization of close-range targets, where high-resolution details are more available. As a result, the $V_d$ method's focus on average distance fails to optimize for the high localization precision required by the mAP50:95 metric.

Using the GELAN-t model, our results show that $V_d$ offers stable and well-balanced performance across different cluster sizes. While it may not always deliver the highest accuracy in every scenario, it consistently avoids the lowest-performing outcomes. For instance, the lowest mAP50 score achieved by $V_d$ was 0.526 in the mid cluster—still higher than the lowest individual viewpoint scores, such as 0.456 from $V(2)$ in the mid cluster, and 0.484 and 0.483 from $V(1)$ and $V(3)$ in the close cluster, respectively.

We further evaluated the distance-based viewpoint selection using the GELAN-ViT-SE model. As shown in Table~\ref{tab:ViT_SE_performance}, the results mirror those from GELAN-t, with the most significant mAP50 improvements observed in the far clusters. While overall mAP50 increased, the mAP50:95 score declined, suggesting improved object detection but reduced precision in bounding box localization under stricter intersection-over-union (IoU) thresholds.

When comparing $V_d$ results from GELAN-t to GELAN-ViT-SE, the latter achieved higher overall scores, with mAP50 increasing from 0.579 to 0.615 and mAP50:95 from 0.187 to 0.190. The most notable improvement was in the close cluster, where mAP50 increased from 0.527 to 0.600 and mAP50:95 from 0.173 to 0.185. In the mid cluster, mAP50 rose from 0.526 to 0.558, while mAP50:95 decreased from 0.170 to 0.166. In the far cluster, mAP50 slightly decreased from 0.712 to 0.708 and mAP50:95 from 0.249 to 0.235. 

Results on the SCS dataset show that the GELAN-ViT-SE-based satellite clustering solution outperforms the GELAN-t-based approach in detection effectiveness. It achieves higher mAP50 scores while operating at lower computational complexity---5.6 giga floating point operations (GFLOPs) versus 7.3 GFLOPs---making it more suitable for real-time onboard deployment.

\subsection{Results Discussion}
Our evaluation results indicate that the proposed satellite clustering solution for SOD outperforms single-satellite approaches based on fixed viewpoints. This solution can also be informed by the performance of the best-performing DL model. Further, while $V_d$ offers valuable benefits, it also presents certain limitations, primarily due to its underlying assumption that ``closer is better.''
While distance to the object is an important factor in SOD, where objects could only occupy a few pixels, our findings suggest that this is not a sufficient condition. The fact that $V_d$ did not outperform the best individual viewpoint in the close and mid clusters indicates that factors beyond proximity, such as occlusions, background complexity, and object orientation, likely play a significant role in SOD.

While the current viewpoint selection strategy is greedy, selecting a single best image rather than combining information across viewpoints, future work could explore other fusion strategies. For instance, score-level fusion would aggregate bounding box confidences across views. Alternatively, a learned policy for viewpoint selection can be trained on scene or image features to dynamically determine the optimal viewpoint rather than relying on static scoring.

\section{Conclusion}
We explored the feasibility of using multiple collaborative satellites to perform SOD tasks, enabling real-time onboard monitoring and enhancing space situational awareness. The proposed solution complements existing spaceborne and ground-based sensing systems while maintaining a low SWaP profile and allowing flexible payload deployment. While this work validates a promising direction for networked, distributed, and embedded AI-based solutions, future efforts are needed to further improve the underlying DL algorithms and the communication mechanisms supporting satellite clustering.


\bibliographystyle{IEEEtran}
\bibliography{references}

\vspace{12pt}

\end{document}